\documentclass{article}
\usepackage{graphics}
\def\abstract#1{\vskip 7mm 
        \begin{center}{\large Abstract}\par \smallskip
                \begin{minipage}[c]{12cm}
                        \small #1
                \end{minipage}
        \end{center}
}
\def\title#1{\begin{center}{\Large\bf #1}\end{center}}
\def\author#1{\vskip 5mm \begin{center}{#1}\end{center}}
\def\address#1{\begin{center}{\it #1}\end{center}}
\makeatletter
\@ifundefined{lesssim}{}{}
\@ifundefined{gtrsim}{}{}
\def\vereq#1#2{\lower3pt\vbox{\baselineskip1.5pt \lineskip1.5pt
\ialign{$\m@th#1\hfill##\hfil$\crcr#2\crcr\sim\crcr}}}
\makeatother

\begin{document}
\begin{flushright}
Guchi-TP-005\\
\end{flushright}
\title{%
  Extremely charged static perfect fluid distributions with dilaton in curved spacetimes
}
\author{%
  Yoshinori Cho,\footnote{E-mail:b2669@sty.cc.yamaguchi-u.ac.jp}
  Yoshitaka Degura\footnote{E-mail:c1997@sty.cc.yamaguchi-u.ac.jp}
  and Kiyoshi Shiraishi\footnote{E-mail:shiraish@sci.yamaguchi-u.ac.jp}
}
\address{%
  Graduate School of Science and Engineering, Yamaguchi University,\\
  Yoshida, Yamaguchi 753-8512, Japan
}
\abstract{
We examine charged static perfect fluid distributions with a dilaton field in the frame-work of general relativity. We consider the case that the Einstein equations reduce to a non-linear version of Poisson equation. We show that Maxwell equation and an equation for a dilaton imply the relation among the charge, mass and dilatonic charge densities.
}

\section{Introduction}
Recently, there has been much interest in the study of Majumdar-Papapetrou metrics~\cite{ref1,ref2,ref3}, discribing the static equilibrium state of extremely charged black holes. For the static Einstein-Maxwell equation with charged dust as the external source of the fields, one can reduce the electrovacuum field equations to the Poisson equation in the flat space~\cite{ref4,ref5,ref6}. In such a system, one can show that the charge and mass densities are equal.

In the low energy limit of string theory, the dilatonic forces as well as gravitational and electric forces are acting among charged matters as long-range forces. In this paper, we study the charged perfect fluid distributions which also couple to a dilaton field in static $(N+1)$ dimensional spacetimes. We find that field equations reduce to a non-linear type of Poisson equation and that Maxwell equation and an equation for a dilaton show the relation among the charge, mass and dilatonic charge densities. We also examine some simple exact solutions.

The organization of this paper is as follows. In the next section, we will show the action and the assumptions on the charged perfect fluid distributions in the static $(N+1)$ dimensional spacetimes, which we consider in the present paper. We reduce the field equations to the non-linear version of Poisson equation in section 3. We find some simple solutions and discuss them in section 4. Finally, section 5 is devoted to conclusion and discussion.
\section{The model}

The action for the fields which mediate long-range forces is
\begin{equation}
 S=\int d^{N+1}x\frac{\sqrt{-g}}{16\pi}\left[R-\frac{4}{N-1}(\nabla\phi)^2 -
e^{-\frac{4a}{N-1}\phi}F^2 \right],
 \label{eq:1}
\end{equation}
where $N$~$(N\geq 3)$ denotes the dimension of space, $R$ is the scalar curvature and $\phi$ is the dilaton field. $F^2 =F^{\mu\nu}F_{\mu\nu}$ and $F_{\mu\nu}$ denotes the Maxwell field strength. The dilaton coupling to the Maxwell term is governed by a constant $a$. The Newton constant is normalized to unity.

Incorporating coupling to matter, we obtain our basic equations:
\begin{equation}
 G_{\mu\nu}-\frac{4}{N-1}\left[\nabla_{\mu}\phi\nabla_{\nu}\phi-\frac{1}{2}g_{\mu\nu}(\nabla\phi)^2 \right]-e^{-\frac{4a}{N-1}\phi}\left[2F_{\mu\nu}^2 -\frac{1}{2}g_{\mu\nu}F^2 \right]=8\pi T_{\mu\nu},
\label{eq:2}
\end{equation}
\begin{equation}
\frac{8}{N-1}\nabla^2 \phi+\frac{4a}{N-1}e^{-\frac{4a}{N-1}\phi}F^2 =
4\pi\frac{8a}{N-1}\rho_{dil},
\label{eq:3}
\end{equation}
\begin{equation}
 \nabla_{\mu}\left[e^{-\frac{4a}{N-1}\phi}F^{\mu\nu}\right]=4\pi j^{\nu},
 \label{eq:4}
\end{equation}
where $G_{\mu\nu}$ is the Einstein tensor. The energy momentum tensor $T_{\mu\nu}$ for a perfect fluid is given by
\begin{equation}
 T_{\mu\nu}=(\rho+p)u_{\mu}u_{\nu}+pg_{\mu\nu},
 \label{eq:5}
\end{equation}
where $\rho$ is the energy density and $u^{\mu}$ is the four velocity. The electric current vector $j^{\mu}$ is defined as
\begin{equation}
 j^{\mu}=\rho_{e}u^{\mu},
 \label{eq:6}
\end{equation}
where $\rho_e$ is the charge density. We have introduced the dilatonic charge density $\rho_{dil}$ in the right hand side of the equation for a dilaton field.


\section{Deriving the non-linear version of Poisson equation}
We assume that the fluid is static and the metric of the static spacetime takes the  form:
\begin{equation}
 ds^2 =-U^{-2}dt^2 +U^{\frac{2}{N-2}}\tilde{g}_{ij}dx^i dx^j ,
 \label{eq:7}
\end{equation}
where $i,j=1,\dots,N$, and both the background metric $\tilde{g}_{ij}$ and $U$ depend only on the space-like coordinates $x^i$.

The Ricci and Einstein tensor components derived from the metric (\ref{eq:7}) are given by
\begin{equation}
 R_{00}=-U^{-2-\frac{2}{N-2}}\tilde{\nabla}_{l}
 \left(\frac{\tilde{\nabla}^{l}U}{U} \right),
\end{equation}
\begin{equation}
G_{ij}=-\frac{N-1}{N-2}\frac{\tilde{\nabla}_{i}U}{U}
      \frac{\tilde{\nabla}_{j}U}{U}+\frac{1}{2}\frac{N-1}{N-2}
      \tilde{g}^{kl}\frac{\tilde{\nabla}_{k}U}{U}
      \frac{\tilde{\nabla}_{l}U}{U}\tilde{g}_{ij}+\tilde{G}_{ij},
\end{equation}
where $\tilde{\nabla}_i$ denotes the $N$ dimensional covariant derivative in terms of $\tilde{g}_{ij}$. $\tilde{G}_{ij}$ is constructed from $\tilde{g}_{ij}$.

Here we should assume that there is only the electric field, namely $F_{0i}\neq 0$ and the others are set to be zero. Then we get
\begin{eqnarray*}
 F^{2}_{00}&=&U^{-\frac{2}{N-2}}\tilde{g}^{kl}F_{0k}F_{0l}, \\
 F^{2}_{ij}&=&-U^{2}F_{0i}F_{0j},  \\
 F^{2}&=&-2U^{2-\frac{2}{N-2}}\tilde{g}^{kl}F_{0k}F_{0l}.
\end{eqnarray*}
In addition, we also put an assumption on the dilatonic field:
\begin{equation}
 e^{-\frac{4a}{N-2}\phi}=U^{2\alpha},
 \label{eq:10}
\end{equation}
where $\alpha$ is a constant.

Then the $(00)$ component in the left hand side of Eq.~(\ref{eq:2}) becomes
\begin{equation}
 R_{00}-e^{-\frac{4a}{N-1}\phi}
 \left[
 2F_{00}^{2}-\frac{1}{N-1}g_{00}F^{2}
 \right]
 =-U^{-2-\frac{2}{N-2}}\tilde{\nabla}^{2}\ln U
 -2U^{2\alpha -\frac{2}{N-2}}\frac{N-2}{N-1}\tilde{g}^{kl}
 F_{0k}F_{0l},
 \label{eq:11}
\end{equation}
while the $(ij)$ component in the left hand side of Eq.~(\ref{eq:2}) is
\begin{eqnarray}
 G_{ij} \! \! \! &-& \! \! \! \frac{4}{N-1}
\left[\nabla_{i}\phi\nabla_{j}\phi-\frac{1}{2}g_{ij}
 (\nabla\phi)^{2}\right]
 -e^{-\frac{4a}{N-1}\phi}\left[2F^{2}_{ij}
 -\frac{1}{2}g_{ij}F^{2}\right] \nonumber 
 \\
 &=& \! \! \! -\frac{N-1}{N-2}\frac{\tilde{\nabla}_{i}U}{U}
    \frac{\tilde{\nabla}_{j}U}{U}-(N-1)\frac{\alpha^{2}}{a^{2}}
    \frac{\tilde{\nabla}_{i}U}{U}\frac{\tilde{\nabla}_{j}U}{U}
    +2U^{2\alpha +2}F_{0i}F_{0j} \nonumber 
 \\
 & & \! \! \! +\frac{1}{2}\frac{N-1}{N-2}\tilde{g}^{kl}
    \frac{\tilde{\nabla}_{k}U}{U}\frac{\tilde{\nabla}_{l}U}{U}
    \tilde{g}_{ij}
    +
    \frac{1}{2}(N-1)\frac{\alpha^{2}}{a^{2}}\tilde{g}^{kl}
    \frac{\tilde{\nabla}_{k}U}{U}
    \frac{\tilde{\nabla}_{l}U}{U}
    \tilde{g}_{ij}
    -
    U^{2\alpha +2}\tilde{g}^{kl}F_{0k}F_{0l}\tilde{g}_{ij}
    \nonumber 
 \\
 & &{} \! \! \!
 +\tilde{G}_{ij}.
 \label{eq:12}
\end{eqnarray}

Here we should suppose that $U^{\alpha +1}=V$, then Eq.~(\ref{eq:11}) is changed into
\begin{eqnarray}
 R_{00}-e^{-\frac{4a}{N-1}\phi}\left[2F^{2}_{00}-\frac{1}{N-1}g_{00}F^{2}\right]
 &=&-U^{-2-\frac{2}{N-2}}\left[\frac{1}{\alpha +1}\frac{\tilde{\nabla}^{2}V}{V}
    -\frac{1}{\alpha +1}\tilde{g}^{kl}
    \frac{\tilde{\nabla}_{k}V}{V}\frac{\tilde{\nabla}_{l}V}{V}
    \right.\nonumber \\
  & & \left.
    +2\frac{N-2}{N-1}V^{2}\tilde{g}^{kl}
    F_{0k}F_{0l}
    \right].
\label{eq:13}
\end{eqnarray}
In order that the second term is canceled by the third one in the right hand side of Eq.~(\ref{eq:13}), we adopt
\begin{equation}
 F_{0k}=\pm\sqrt{\frac{N-1}{2(N-2)}}
 \sqrt{\frac{1}{\alpha +1}}
 \frac{\tilde{\nabla}_{k}V}{V^{2}}.
 \label{eq:14}
\end{equation}
On the other hand, assuming Eq.~(\ref{eq:14}), we can reduce Eq.~(\ref{eq:12}) to
\begin{eqnarray}
 G_{ij} \! \! \! &-& \frac{4}{N-1}\left[
 \nabla_{i}\phi\nabla_{j}\phi-\frac{1}{2}g_{ij}(\nabla\phi)^{2}\right]
 -e^{-\frac{4a}{N-1}\phi}\left[2F^{2}_{ij}
 -\frac{1}{2}g_{ij}F^{2}\right]
 \nonumber 
 \\
 &=& \! \! \! \left(
 -\frac{1}{N-2}\frac{1}{\alpha +1}-\frac{\alpha^{2}}{a^{2}}
  \frac{1}{\alpha +1}+\frac{1}{N-2}
 \right)\frac{N-1}{\alpha +1}
 \frac{\tilde{\nabla}_{i}U}{U}\frac{\tilde{\nabla}_{j}U}{U}
 \nonumber  
 \\
 & & \! \! \! 
 -\frac{1}{2}
 \left(
 -\frac{1}{N-2}\frac{1}{\alpha +1}-\frac{\alpha^{2}}{a^2}
 \frac{1}{\alpha +1}+\frac{1}{N-2}
 \right)\frac{N-1}{\alpha +1}
 \tilde{g}^{kl}
 \frac{\tilde{\nabla}_{k}U}{U}\frac{\tilde{\nabla}_{l}U}{U}
 \tilde{g}_{ij}
 \nonumber 
 \\ 
 & & \! \! \! +\tilde{G}_{ij}.
 \label{eq:15}
\end{eqnarray}
In order to eliminate the first and second terms in the right hand side of Eq.~(\ref{eq:15}), we take
\begin{equation}
 \alpha =\frac{a^2}{N-2}.
\end{equation}

Consequently, we reduce the left hand side of Eq.~(\ref{eq:2}) to the following equations:
\begin{equation}
 R_{00}-e^{-\frac{4a}{N-1}\phi}\left[
 2F^{2}_{00}-\frac{1}{N-1}g_{00}F^{2}
 \right]
 =
 -U^{-2-\frac{2}{N-2}}\frac{N-2}{N-2+a^{2}}
 \frac{1}{V}\tilde{\nabla}^{2}V,
 \label{eq:17}
\end{equation}
\begin{equation}
 G_{ij}-\frac{4}{N-1}
 \left[
 \nabla_{i}\phi\nabla_{j}\phi-\frac{1}{2}g_{ij}
 (\nabla\phi)^{2}
 \right]
 -e^{-\frac{4a}{N-1}\phi}
 \left[2F^{2}_{ij}
 -\frac{1}{2}g_{ij}F^{2}
 \right]
 =\tilde{G}_{ij}.
 \label{eq:18}
\end{equation}
We should remember that
\begin{equation}
 e^{-\frac{4a}{N-1}\phi}=U^{\frac{2a^{2}}{N-2}}
 =V^{\frac{2a^{2}}{N-2+a^{2}}},
 \label{eq:19}
\end{equation}
\begin{equation}
 F_{0k}=\pm\sqrt{\frac{N-1}{2(N-2+a^{2})}}
 \frac{\tilde{\nabla}_{k}V}{V^{2}}.
 \label{eq:20}
\end{equation}

Finally, using Eqs.~(\ref{eq:5}),~(\ref{eq:6}) and~(\ref{eq:17}-\ref{eq:20}), we reduce the field equations~(\ref{eq:2}),~(\ref{eq:3})~and~(\ref{eq:4})~simply to the following equations:
\begin{equation}
 \tilde{\nabla}^2 V+8\pi\frac{N-2+a^2}{N-1}
 V^{\frac{N+a^2}{N-2+a^2}}(\rho+\frac{N}{N-2}p)=0 \ ,
 \label{eq:21}
\end{equation}
\begin{equation}
 \tilde{R}_{ij}=
 -\frac{16\pi p}{N-2}
  V^{\frac{2}{N-2+a^2}}\tilde{g}_{ij} \ ,
\end{equation}
\begin{equation}
 \rho_{dil}=\rho+\frac{N}{N-2}p \ ,
\end{equation}
\begin{equation}
 \rho_e=\pm e^{-\frac{2a}{N-1}\phi}\sqrt{\frac{2(N-2+a^2)}{N-1}}
 \left(\rho +\frac{N}{N-2}p \right) \ .
 \label{eq:24}
\end{equation}
Therefore, these equations represent the Einstein, Maxwell and dilaton equations.

Here we think about Eq.~(\ref{eq:24}) for the dust case $(p=0)$. The action for  particles, of which coordinates are denoted by $x^{\mu}$, can be written as:
\begin{equation}
 I=-\sum_{a}\int ds_{a}\left[m_a e^{\frac{2a}{N-1}\phi}+e_a A_{\nu}\frac{dx^{\nu}_{a}}{ds_a} \right],
\end{equation}
where $m_a$ and $e_a$ stand for the mass and electric charges of the particles. Suppose that the distribution of these particles represents the matter densities. One can find that the dilatonic charge density is proportional to the charge density. Thus, we can recognize that the relationship between the charge density and the mass density is $\rho_e \propto \pm e^{-\frac{2a}{N-1}\phi}\rho$, because each electric charge $e_a$ is a constant. In the next section, we discuss the some explicit solutions of Eq.~(\ref{eq:21}).


\section{Exact solutions}
For the dust case $(p=0)$, we find some simple exact solutions of Eq.~(\ref{eq:21}), which do not have the singularities. When spherical symmetry is assumed, the non-linear version of Poisson equation takes the following form:
\begin{equation}
 \frac{d^2 V}{dr^2}+\frac{N-1}{r}\frac{dV}{dr}+
 8\pi\rho\frac{N-2+a^2}{N-1}V^{\frac{N+a^2}{N-2+a^2}}=0.
 \label{eq:26}
\end{equation}

If we put the following condition on the energy density:
\begin{equation}
 \rho=\frac{A}{8\pi}\frac{N-1}{N-2+a^2}V^{-\frac{N+a^2}{N-2+a^2}},
 \label{eq:27}
\end{equation}
we can find that the solution is 
\begin{equation}
 V(r)=B-\frac{Ar^2}{6N}.
\end{equation}
Here $A$ and $B$ are constants.

We show that the energy density $\rho$ for a certain value of total mass plotted against $r$ for $a^2 =0$, $a^2 =\frac{N-1}{2N}$, $a^2 =1$ and $a^2 =N$ in Fig.~1(a) in the case of $N=3$. In Fig.~1(b), $1-U^2$ is plotted against $r$ for the same coupling constants. Here the energy density is matched to the one for the vacuum solution at $r=2$. Fig.~2 is drawn with the same conditions of Fig.~1, except for $N=5$ and Fig.~3 is also, except for $N=9$.

If we put another condition:
\begin{equation}
 \rho=\frac{C^2}{8\pi}\frac{N-1}{N-2+a^2}V^{-\frac{2}{N-2+a^2}},
 \label{eq:29}
\end{equation}
the solution is 
\begin{equation}
V(r)=D\frac{J_{(N-2)/2}(Cr)}{r^{(N-2)/2}}.
\end{equation}
Here $C$ and $D$ are constants, and $J_{\nu}(z)$ is the Bessel function. If we choose $N=3$ and $a^2 =0$, then we can obtain the same results of G\"{u}rses~\cite{ref4}.

We show that the energy density $\rho$ for a certain value of total mass plotted against $r$ for $a^2 =0$, $a^2 =\frac{N-1}{2N}$, $a^2 =1$ and $a^2 =N$ in Fig.~4(a) in the case of $N=3$. In Fig.~4(b), $1-U^2$ is plotted against $r$ for the same coupling constants. Here the energy density is matched to the one for the vacuum solution at $r=2$. Fig.~5 is drawn with the same conditions of Fig.~4, except for $N=5$ and Fig.~6 is also, except for $N=9$.

Varela considered the case that Eq.~(\ref{eq:26}) can be reduced to the sine-Gordon equation~\cite{ref6}. Using the new radial coodinate $\tau=\frac{1}{r^{N-2}}$ to rearrange Eq.~(\ref{eq:26}), we obtain:
\begin{equation}
 \frac{d^2 V}{d\tau^2}+8\pi\rho\frac{N-2+a^2}{(N-2)^2 (N-1)}
 \tau^{-\frac{2(N-1)}{N-2}}V^{\frac{N+a^2}{N-2+a^2}}=0.
 \label{eq:31}
\end{equation}
If we assume
\begin{equation}
 \rho=\frac{E^2}{8\pi}\frac{(N-2)^2 (N-1)}{N-2+a^2}\tau^{\frac{2(N-1)}{N-2}}
 (\sin V) V^{-\frac{N+a^2}{N-2+a^2}},
 \label{eq:32}
\end{equation}
then, Eq.~(\ref{eq:31}) reduces to the sine-Gordon equation
\begin{equation}
 \frac{d^2 V}{d\tau^2}+E^2 \sin V=0,
\end{equation}
which has the solutions
\begin{equation}
 V(\tau)=2\arcsin[\tanh(E\tau+F)]+2n\pi \ ,
 \label{eq:34}
\end{equation}
where $n$ is an arbitrary integer, $F$ is an integration constant, and $E$ is assumed to be positive. We consider only the case $n=0$. If we choose the integration constant $F$ for
\begin{equation}
 F=\frac{1}{2}\ln \left[\frac{1+\sin(1/2)}{1-\sin(1/2)}\right],
 \label{eq:35}
\end{equation}
then the spacetime corresponding to Eq.~(\ref{eq:34}) and Eq.~(\ref{eq:35}) becomes asmptotically flat~\cite{ref6}.

We show that the energy density $\rho$ for a certain value of total mass plotted against $r$ for $a^2 =0$, $a^2 =\frac{N-1}{2N}$, $a^2 =1$ and $a^2 =N$ in Fig.~7(a) in the case of $N=3$. In Fig.~7(b), $1-U^2$ is plotted against $r$ for the same coupling constants.

Here, in these figures, we find that the energy density decreases as the coupling constant $a^2$ increases. We also find that the difference between the energy densities gets narrow for the various values of $a^2$ and the contrast (i.e., the difference between the energy density at $r=0$ and the one at $r=2$) decreases as the dimension of space $N$ increases.


\section{Conclusion and discussion}
In this paper, we have investigated charged static perfect fluid distributions with the dilaton field in the frame-work of general relativity. As shown in section 3, the Einstein equations have reduced to the non-linear version of Poisson equation, and the Maxwell equation and the equation for the dilaton have implied the relation among the charge, mass and dilatonic charged densities. For the dust case, one can find that the relationship between the charge density and the mass density is $\rho_e \propto \pm e^{-\frac{2a}{N-1}\phi}\rho$, because, for point particles, the dilaton does not couple to the electric charge but to the mass.

In section 4, we have found simple exact solutions of Eq.~(\ref{eq:21}) corresponding to certain energy densities. We have found that the energy density decreases as the coupling constant $a^2$ increases. We have found that the energy density decreases as the coupling constant $a^2$ increases. We have found that the difference between the energy densities gets narrow for the various values of $a^2$ and the contrast (i.e., the difference between the energy density at $r=0$ and the one at $r=2$) decreases as the dimension of space $N$ increases.

We have not yet dealt with Eq.~(\ref{eq:21}) on the condition for $p\neq 0$. Recently, Ida found some exact charged solutions in this situation~\cite{ref5}. We will study the non-zero pressure case with a dilaton field in $(N+1)$ dimensions. Also we have not yet considered the case of $N=2$, which we have only thought of the equilibrium between the dilatonic attractions and the electric repulsions. We must continue to make every effort to study these situations.

\begin{figure}
 \includegraphics{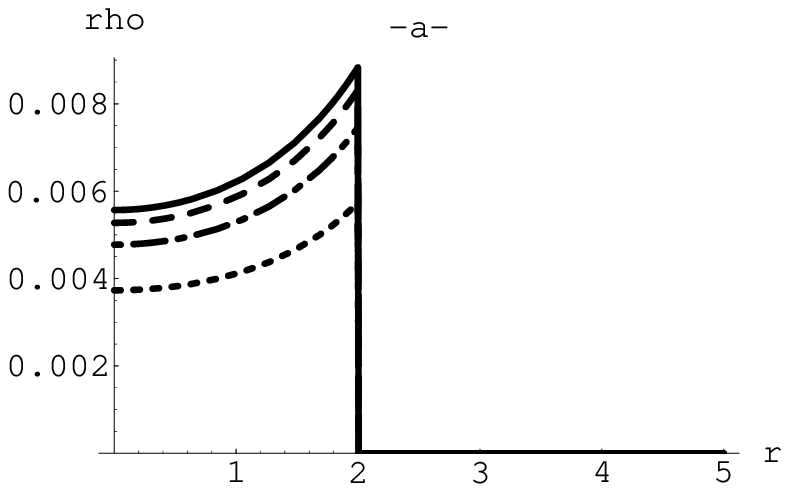}
 \includegraphics{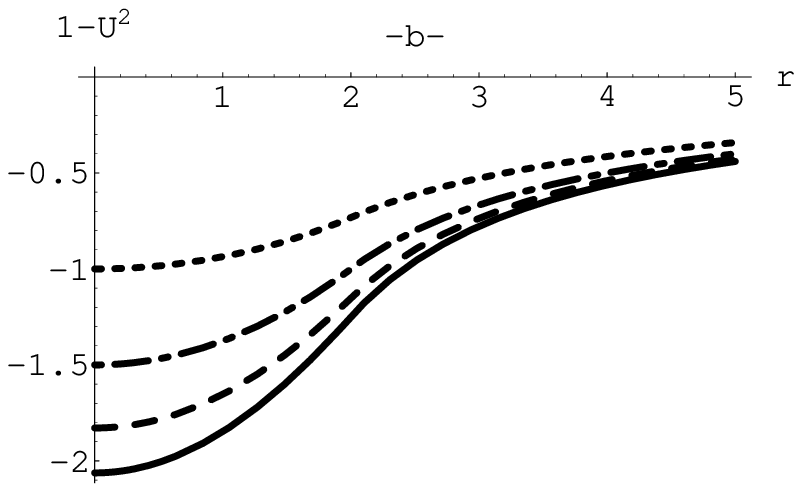}
 \caption{(a) the energy density $\rho$ of Eq.~(\ref{eq:27}) is plotted against $r$ in the case of $N=3$. (b) $1-U^2$ is plotted against $r$ for the same coupling constants. Here the energy density is matched to the one for the vacuum solution at $r=2$. The solid line corresponds to $a^2 =0$, the dashed line corresponds to $a^2 =1/3$, the dot-dashed line corresponds to $a^2 =1$, the dotted line corresponds to $a^2 =3$.}
\end{figure}

\begin{figure}
 \includegraphics{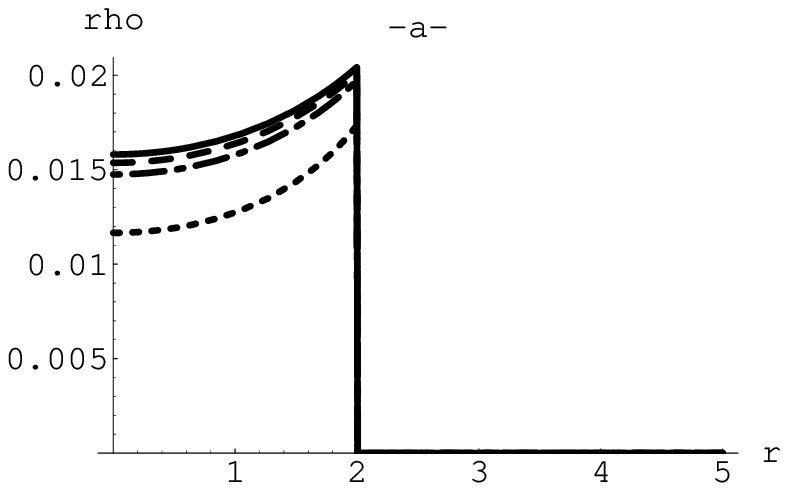}
 \includegraphics{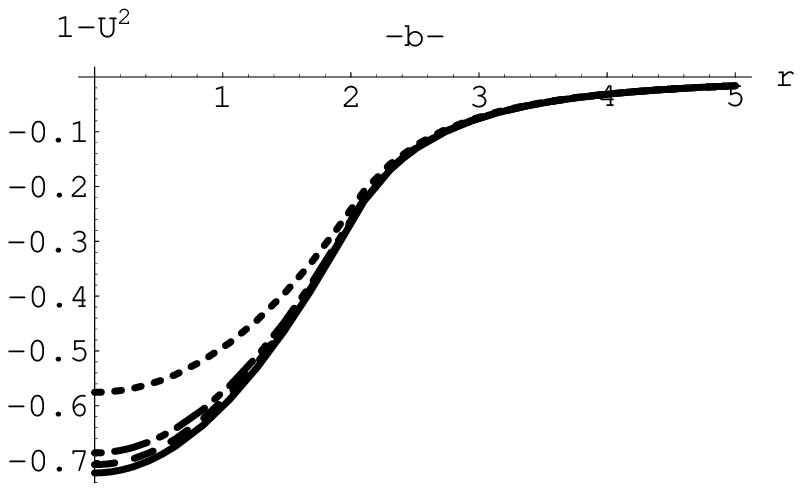}
 \caption{(a) the energy density $\rho$ of Eq.~(\ref{eq:27}) is plotted against $r$ in the case of $N=5$. (b) $1-U^2$ is plotted against $r$ for the same coupling constants. Here the energy density is matched to the one for the vacuum solution at $r=2$. The solid line corresponds to $a^2 =0$, the dashed line corresponds to $a^2 =2/5$, the dot-dashed line corresponds to $a^2 =1$, the dotted line corresponds to $a^2 =5$.}
 \label{(a)}
\end{figure}

\begin{figure}
 \includegraphics{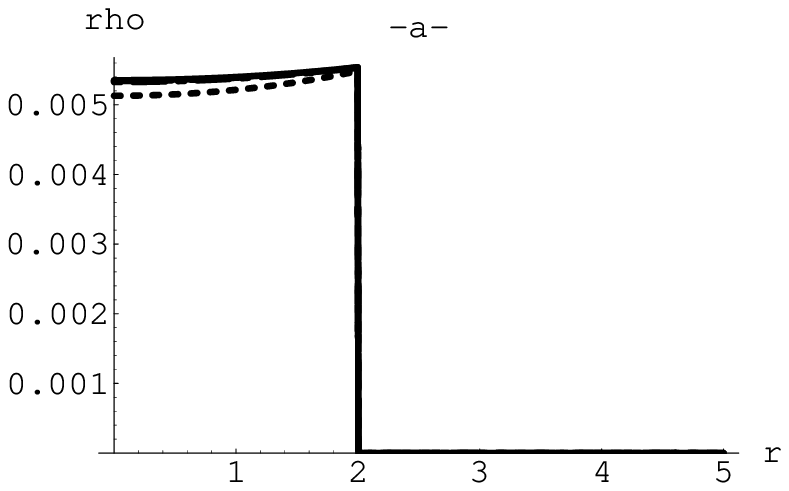}
 \includegraphics{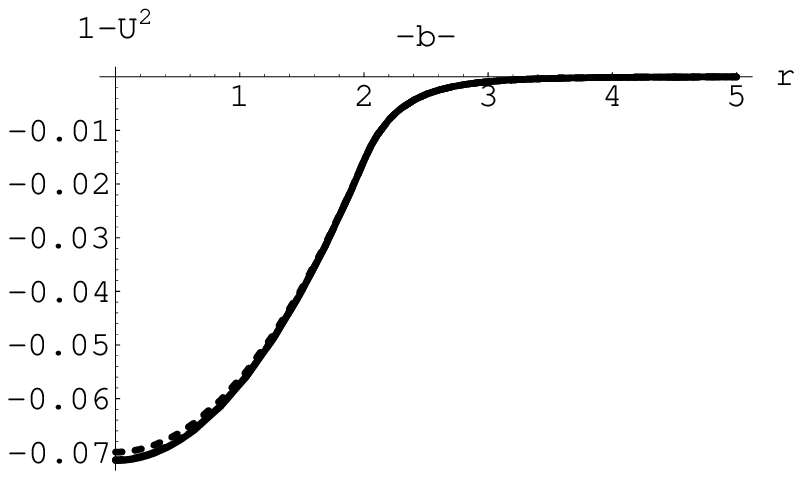}
 \caption{(a) the energy density $\rho$ of Eq.~(\ref{eq:27}) is plotted against $r$ in the case of $N=9$. (b) $1-U^2$ is plotted against $r$ for the same coupling constants. Here the energy density is matched to the one for the vacuum solution at $r=2$. The solid line corresponds to $a^2 =0$, the dashed line corresponds to $a^2 =4/9$, the dot-dashed line corresponds to $a^2 =1$, the dotted line corresponds to $a^2 =9$.}
\end{figure}

\begin{figure}
 \includegraphics{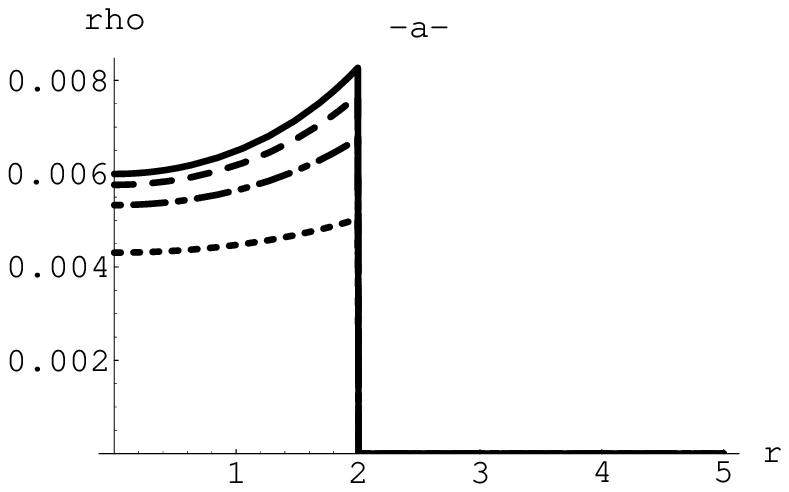}
 \includegraphics{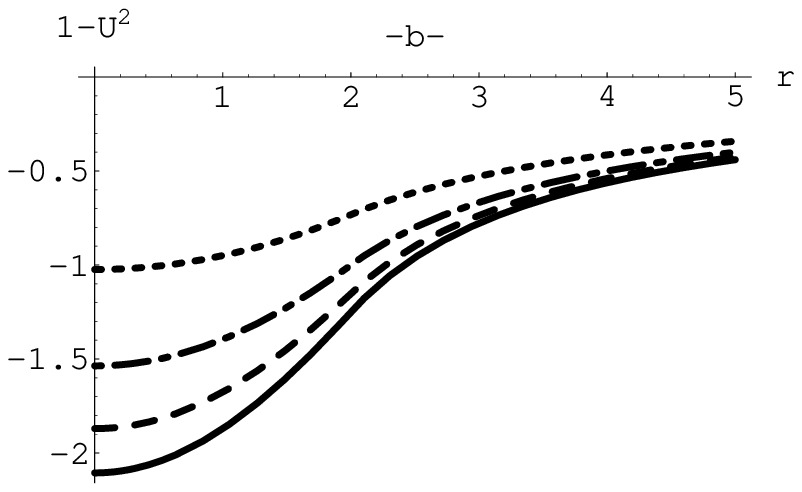}
 \caption{(a) the energy density $\rho$ of Eq.~(\ref{eq:29}) is plotted against $r$ in the case of $N=3$. (b) $1-U^2$ is plotted against $r$ for the same coupling constants. Here the energy density is matched to the one for the vacuum solution at $r=2$. The solid line corresponds to $a^2 =0$, the dashed line corresponds to $a^2 =1/3$, the dot-dashed line corresponds to $a^2 =1$, the dotted line corresponds to $a^2 =3$.}
\end{figure}

\begin{figure}
 \includegraphics{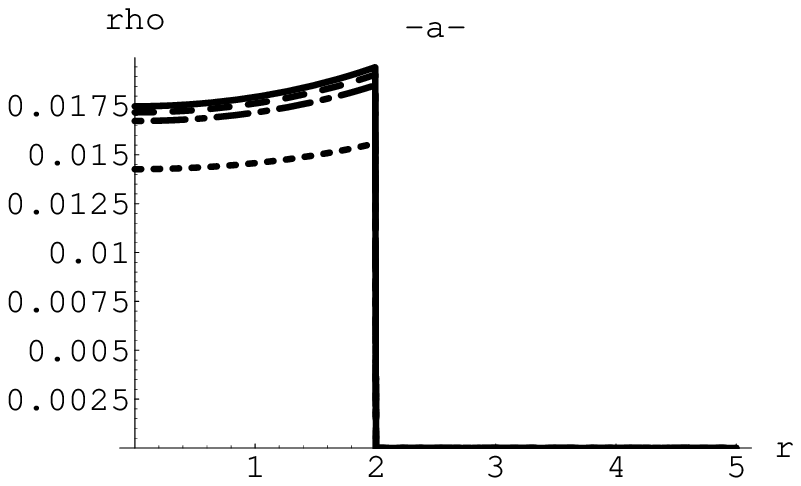}
 \includegraphics{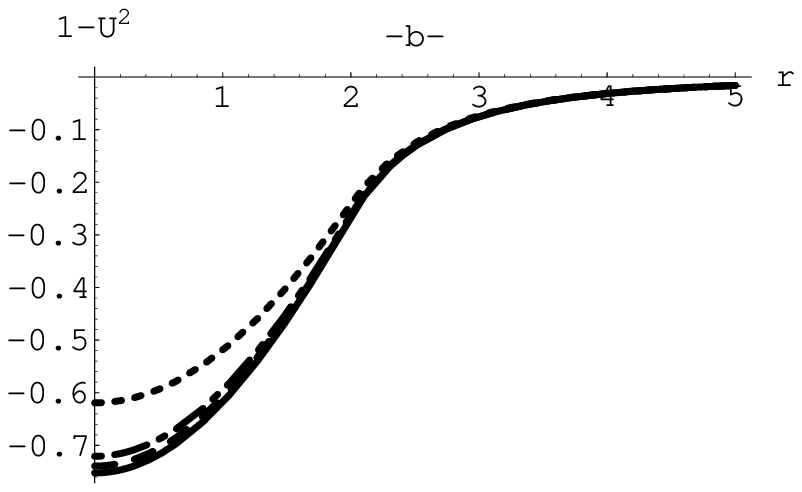}
 \caption{(a) the energy density $\rho$ of Eq.~(\ref{eq:29}) is plotted against $r$ in the case of $N=5$. (b) $1-U^2$ is plotted against $r$ for the same coupling constants. Here the energy density is matched to the one for the vacuum solution at $r=2$. The solid line corresponds to $a^2 =0$, the dashed line corresponds to $a^2 =2/5$, the dot-dashed line corresponds to $a^2 =1$, the dotted line corresponds to $a^2 =5$.}
\end{figure}

\begin{figure}
 \includegraphics{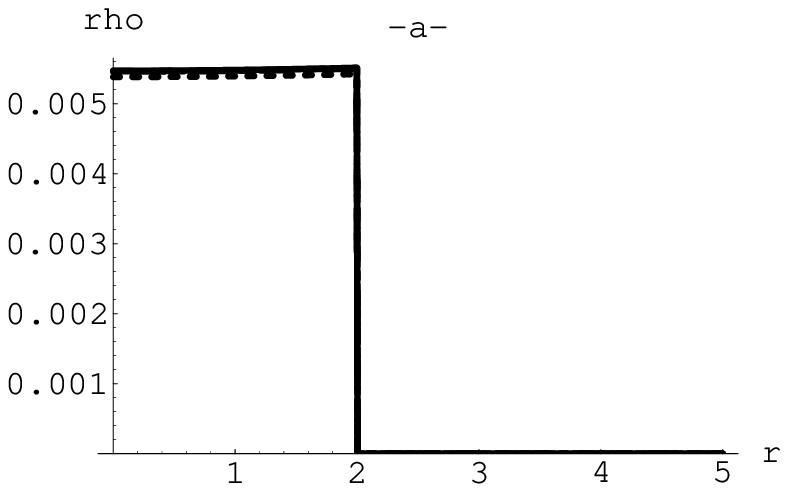}
 \includegraphics{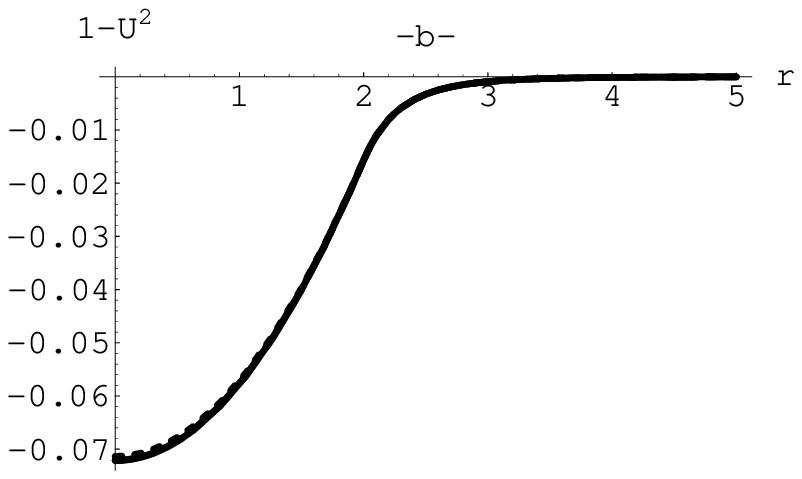}
 \caption{(a) the energy density $\rho$ of Eq.~(\ref{eq:29}) is plotted against $r$ in the case of $N=9$. (b) $1-U^2$ is plotted against $r$ for the same coupling constants. Here the energy density is matched to the one for the vacuum solution at $r=2$. The solid line corresponds to $a^2 =0$, the dashed line corresponds to $a^2 =4/9$, the dot-dashed line corresponds to $a^2 =1$, the dotted line corresponds to $a^2 =9$.}
\end{figure}

\begin{figure}
 \includegraphics{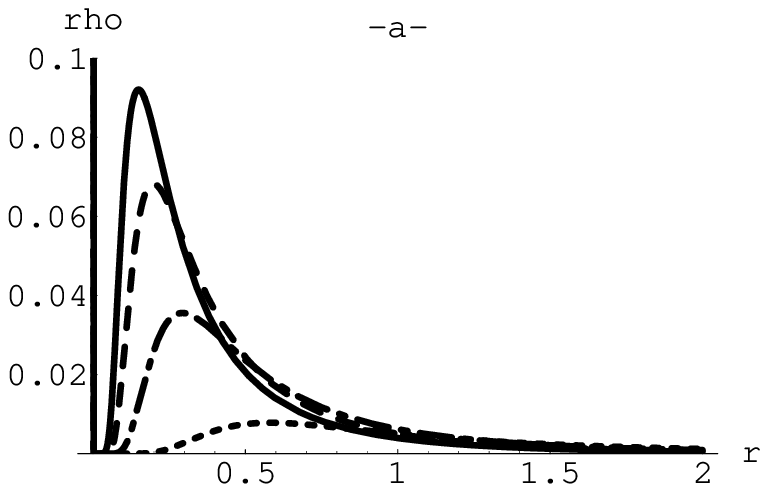}
 \includegraphics{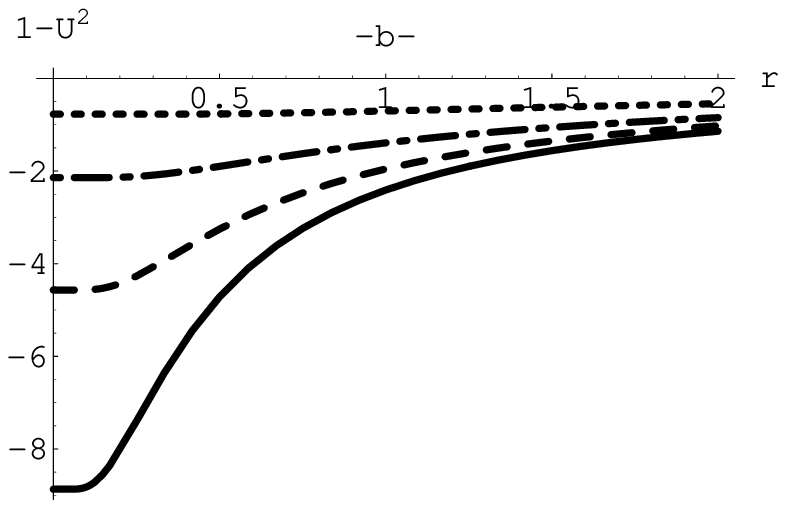}
 \caption{(a) the energy density $\rho$ of Eq.~(\ref{eq:32}) is plotted against $r$ in the case of $N=3$. (b) $1-U^2$ is plotted against $r$ for the same coupling constants. The solid line corresponds to $a^2 =0$, the dashed line corresponds to $a^2 =1/3$, the dot-dashed line corresponds to $a^2 =1$, the dotted line corresponds to $a^2 =3$.}
\end{figure}

\section*{Acknowledgement}
The authors would like to thank N.~Kan, M.~Ooho and T.~Watabe for useful advice and support.


\end{document}